\documentclass[twocolumn,10pt]{article}
\usepackage[utf8]{inputenc}
\usepackage[dvips]{graphicx}
\usepackage{amsthm}
\usepackage{amsmath}
\usepackage{amssymb}
\usepackage{mathrsfs} 
\usepackage{natbib}
\usepackage{enumerate}
\usepackage[dvips]{graphicx}
\usepackage[usenames,dvipsnames]{xcolor}

\newcommand{\ud}{\mathrm{d}}
\newcommand{\ue}{\mathrm{e}}

\title{Second-order dust perturbations of the non-flat FLRW model with the positive cosmological constant.}
\author{Szymon Sikora \\
	\scriptsize{\textit{Astronomical Observatory, Jagiellonian University, Orla 171, 30-244 Kraków, Poland}} }

\begin{document}
	
	\twocolumn[
	\begin{@twocolumnfalse}
		\maketitle
		\begin{abstract}
			In this paper, a specific solution to the second-order cosmological perturbation theory is given. Perturbations are performed around any FLRW spacetime filled with dust and with a positive cosmological constant. In particular, with a possibly non-vanishing spatial curvature. The adopted symmetry condition allows us to simplify the equations, leaving us with a great deal of freedom to choose the density distribution. In the result, we get a relatively simple metric of an inhomogeneous cosmological model, which will give a perfect tool for studying the influence of the local inhomogeneities onto the cosmological observables. 
		\end{abstract}
		\vspace{1cm}
	\end{@twocolumnfalse}
	]
	
	\small
	\section{Introduction}
	It is not possible to imagine cosmology without the Friedmann-Lema\^itre-Robertson-Walker (FLRW) model describing a universe
	which is spatially homogeneous and isotropic. However, apart from a significant success in explaining variety of the cosmological observables by the FLRW model, there are some issues which remain to be clarified. 
	
	In recent years, the precision of the measurements of the Lema\^itre-Hubble constant $H_0$ has increased. Unfortunately, the estimations of its value using different methods have become inconsistent. The CMB observations provide the value $H_0=67.37 \pm 0.54\,\mathrm{km/s/Mpc}$ \cite{2020A&A...641A...6P} lower than the estimates based on standard candles $H_0=74.03 \pm 1.42\,\mathrm{km/s/Mpc}$ \cite{2019ApJ...876...85R}. This discrepancy is known as the Hubble tension problem. For more details one can see a review \cite{2021CQGra..38o3001D} and the references therein. Among many possible solutions, the influence onto the local $H_0$ measurements by the inhomogeneity around the observer is reliable and based on standard physics. The idea that the matter inhomogeneities on a different scales could affect the cosmological observations is not new and appear also in earlier works, e. g. \cite{2008PhRvD..78h3531L}. In the context of the Hubble tension, a directional measurements of the Lema\^itre-Hubble constant \cite{2016JCAP...04..036B} are important. A theoretical analysis of this phenomenon within the Szekeres solution can be found in \cite{2016JCAP...06..035B}. We can mention also a recent attempt in explaining quantitatively Hubble parameter inconsistencies within the inhomogeneous cosmology \cite{2020CQGra..37p4001H}.
	
	Another issue is the problem of the averaging. It is believed that the homogeneous density of the FLRW model is an effect of the averaging of the matter inhomogeneities over a large scales. However, calculation of the averages in a curved spacetime is highly non-trivial. It could happen that the averaged expansion of the inhomogenous universe differs from the expansion of the corresponding FLRW model which density equals the averaged density of the inhomogeneous model. This effect is known as the cosmological backreaction. There exist several averaging schemes which lead to a different conclusions about the importance of the backreaction effect. Within the Green-Wald averaging scheme the authors conclude that the backreaction effect is negligible \cite{2011PhRvD..83h4020G,2016CQGra..33l5027G}, while in the Buchert framework some papers suggest that the backreaction could be important \cite{2000GReGr..32..105B,2015CQGra..32u5021B,2018MNRAS.473L..46B,2018CQGra..35xLT02B} or even it could explain the accelerated expansion of the universe without the need of the cosmological constant \cite{2013JCAP...10..043R}. Recently, other approaches to the averaging problem are investigated. An interesting example is the analysis of some observables and averages over the past light cone \cite{2021LMaPh.111...53C,2022arXiv220210798B,2011JCAP...07..008G}.
	
	To study the effects caused by the inhomogeneities one can analyze exact solutions to the Einstein equations, which are generalizations of the FLRW spacetime. The most important of such models are the Lema\^itre-Tolman-Bondi spacetime \cite{1947MNRAS.107..410B} and the Szekeres solution \cite{1975CMaPh..41...55S}. It is also possible to immerse several substructures described by these solutions into the homogeneous background within the Swiss-cheese framework \cite{2008JCAP...06..021B,2013PhRvD..87l3526F,2013JCAP...12..051L,2015JCAP...10..057L,2021PhRvD.104d3505K}. Although the Einstein equations are very complicated and it is hard to discover new solutions, there is some progress in this field, e. g. \cite{2021CQGra..38a5016N} found a new class of interesting models. Another possibility is to study the evolution of the inhomogeneities within the numerical relativity \cite{2017PhRvD..95f4028M,2016JCAP...07..053A,2014CQGra..31w4006A,2022PhLB..82636911E} or by using the perturbative methods like the post-Newtonian formalism \cite{2015PhRvD..91j3532S,2016PhRvD..93h9903S,2017JCAP...07..028S}.
	
	In this paper we focus on the cosmological perturbation theory. Many years after original papers by Lifshitz \cite{1946ZhETF..16..587L,2017GReGr..49...18L} and Bardeen \cite{1980PhRvD..22.1882B} there is still progress in this area. In particular, there are attemts to analyze the perturbations beyond the linear order \cite{2014JCAP...07..037N,2017JCAP...10..027G,2017PhRvD..96j3522W}. In \cite{2009JCAP...11..012C} the third-order perturbations are considered. The matter inhomogeneities in the form of the ensemble of the point masses are studied up to second order in \cite{2017ApJ...845..153B}. There are also investigations of particular gauge conditions \cite{2006PhRvD..73d4021H} and a complementary approach regarding gauge-independent variables \cite{2005AcPPB..36.2133G}.
	
	The perturbative equations in the second order become quite complicated. For that reason, instead of studying the perturbation theory in general, we are looking for a specific solution for which the resulting metric tensor is as simple as possible. In \cite{2017PhRvD..95f3517S} we found interesting solution within the linear perturbation theory, where the matter inhomogeneities form an infinite periodic cubic lattice on the spatially flat Einstein-de Sitter background. A periodic lattice is a natural choice for a numerical relativity and it is present also in the other works, e. g. in the post-Newtonian formalism \cite{2017JCAP...07..028S}. In \cite{2019PhRvD..99h3521S} we generalize the presented solution up to the third-order perturbations and analyze the light propagation through this inhomogeneous model. In both papers \cite{2017PhRvD..95f3517S,2019PhRvD..99h3521S} a decaying mode was considered. The subsequent paper \cite{2021EPJC...81..208S} was dedicated to the analysis of the growing mode and the fourth-order perturbations were taken into account. Since a high order perturbations were present, a relatively large amplitude of the inhomogeneities was admissible. The papers \cite{2017PhRvD..95f3517S,2019PhRvD..99h3521S,2021EPJC...81..208S} considered a spatially flat background without the cosmological constant. The main purpose of the present paper is to generalize these results such that the considered background could be any FLRW cosmological model filled with dust, with possibly non-vanishing spatial curvature, and with a positive cosmological constant. For that purpose, we consider a two-parameter perturbation theory, where the curvature parameter $k$ is treated as the second perturbative variable. The idea of a two-parameter perturbations is not new and appear in different contexts, e. g. \cite{2017PhRvD..95d3503G,2017PhRvD..96j3508G,2021JCAP...08..048G}.
	
	The paper is organized as follows. In Sec. \ref{sec:construction} we present the main idea of the currect work. The specific solution to the perturbation theory in the first and second order is presented in Sec. \ref{Sec:1stOrder} and Sec. \ref{Sec:2ndOrder} respectively. Then, the conclusions are given.
	
	\section{The model construction.}\label{sec:construction}
	\subsection{The metric.}
	Let us consider Cartesian-like coordinates \mbox{$(t,x,y,z)$}, in which the line element reads: 
	\begin{equation}\label{eqn:metric_GaussianNormal}
		\ud s^2=-\ud t^2+a(t)^2\,c_{i\:\!j}(\lambda,k,x^\mu)\,\ud x^i\,\ud x^j\,.
	\end{equation}
	For the metric written in this way, \mbox{$(t,x,y,z)$} are the Gaussian normal coordinates. It is always possible to transform the metric into the form (\ref{eqn:metric_GaussianNormal}) locally, but it is not obvious whether one could do it for the whole space-time. The vector field $(U^\mu)=(1,0,0,0)$ is tangent to the time-like geodesic at each point. Due to the attractive nature of gravity, the geodesics could be focused and eventually cross with each other. In that case, the Gaussian normal coordinates are not suitable because the point where two geodesics meet would have two distinct coordinate values. However, this is not the case for the model presented in this article. As we shall see in Sect. \ref{Sec: gauge}, the expansion scalar $\theta=\nabla_\mu\,U^\mu$ is always positive here. Since space is expanding everywhere, geodesics which start at different points on some hypersurface of constant time never cross in the future. In effect, the Gaussian normal coordinates are properly defined in the whole spacetime. Due to the positive values of $\theta$, the model does not intend to describe the formation of structures, but provides a tool for handling the overall differences between cosmic voids and regions populated with galaxy clusters.  
	
	The metric functions $c_{i\:\!j}$ depend on the coordinates and two parameters with the following meaning. For $\lambda$ equal to zero, the model should reduce to the FLRW spacetime with arbitrary curvature parameter $k$:
	\begin{equation}\label{eqn:background1}
		\lim\limits_{\lambda\rightarrow 0} c_{i\:\!j}=\frac{\delta_{i\:\!j}}{\left(1+\frac{1}{4}k\,(x^2+y^2+z^2) \right)^2} \equiv \widetilde{c}_{i\:\!j} \,.
	\end{equation}
	This means that $a(t)$ is the scale factor of the background. The energy-momentum tensor of matter that fills spacetime is $T^\mu\:\!{}_\nu=\rho\,U^\mu\,U_\nu$, where $(U^\mu)=(1,0,0,0)$. For $\lambda\neq 0$, the density $\rho$ could be inhomogeneous and the parameter $\lambda$ is proportional to the amplitude of the inhomogeneities.
	
	\subsection{The perturbations.}
	It is difficult to find the functions $c_{i\:\!j}$ that meet the conditions given above. To deal with this problem, we will consider some simplifications. Let us assume that we can express metric functions as a power series:
	\begin{equation}\label{eqn:metric_c}
		c_{i\:\!j}(\lambda,k,x^\mu)=\sum\limits_{l=0}^N\sum\limits_{m=0}^{l}c_{i\:\!j}^{(l-m,m)}(x^\mu)\,\lambda^{l-m}\,k^m\,,
	\end{equation}
	where the functions $c_{i\:\!j}^{(l,m)}$ in each order, labeled with the indices $(l,m)$, depend on the coordinates only. Then, the metric is differentiable in parameters $(\lambda,k)$, and it is possible to calculate the two-dimensional Taylor expansion of the Einstein tensor around $(\lambda,k)=(0,0)$. I denote the two-dimensional Taylor series of any function $f(\lambda,k)$, truncated at the $N$-th order by:
	\begin{equation}
		\mathcal{T}_N[f(\lambda,k)]=\sum\limits_{n=0}^{N}\frac{1}{n!}\sum\limits_{m=0}^{n}\binom{n}{m}\frac{\partial^n f}{\partial\lambda^{n-m}\partial k^m}\Bigg |_{\substack{\lambda=0\\k=0}} \lambda^{n-m}\,k^m\,.
	\end{equation} 
	The Einstein equations hold approximately:
	\begin{equation}\label{eqn:Einstein}
		\mathcal{T}_N[G^\mu\:\!{}_\nu(\lambda,k)]=8\pi\,\rho\,U^\mu\,U_\nu-\Lambda\,\delta^\mu_\nu+8\pi\Delta T^\mu\:\!{}_\nu\,,
	\end{equation}
	since the truncation of the Taylor series of Einstein tensor elements introduces some uncertainty $\Delta T^\mu\:\!{}_\nu$. \mbox{We demand} that the four-velocity $(U^\mu)=(1,0,0,0)$ is unperturbed; the density $\rho(t,\vec{x})$ is a known specified function and the uncertainty $|\Delta T^\mu\:\!{}_\nu|\ll 1$ is small. For exact solutions to the Einstein equations, one would expect $G^\mu\:\!{}_\nu$ instead of $\mathcal{T}_N[G^\mu\:\!{}_\nu]$ on the left-hand side of (\ref{eqn:Einstein}) and $\Delta T^\mu\:\!{}_\nu=0$. However, in the real Universe the pressure of the intra-cluster gas and the proper motions of the galaxy cluster members introduce departures from the energy-momentum tensor of the dust. In the previous paper \cite{2021EPJC...81..208S}, we estimated the upper bound of the uncertainty $|\Delta T^\mu\:\!{}_\nu|\approx 10^{-6}\,\rho$ consistent with a typical value of the velocity dispersion of galaxies observed in galaxy clusters. In the matter-dominated era, it is then justified to describe the matter content of the universe with the approximate energy-momentum tensor $T^\mu\:\!{}_\nu=\rho\,U^\mu\,U_\nu+\Delta T^\mu\:\!{}_\nu$ satisfying this empirical constraint.
	
	\subsection{The background.}
	The second simplification concerns the background. The form of the metric functions (\ref{eqn:metric_c}) is not compatible with the limit (\ref{eqn:background1}). For that reason, we consider a weaker condition:
	\begin{equation}\label{eqn:background2}
		\lim\limits_{\lambda\rightarrow 0} c_{i\:\!j}=\mathcal{T}_N[\widetilde{c}_{i\:\!j}] \,.
	\end{equation}
	Because the parameter $k$ is small, the difference between $\widetilde{c}_{i\:\!j}$ and $\mathcal{T}_N[\widetilde{c}_{i\:\!j}]$ remains small either, if only the coordinate distance $r=\sqrt{x^2+y^2+z^2}$ is not too large. For that reason, the model presented in this paper can only describe the local neighborhood of the chosen observer. The form of the metric functions $c_{i\:\!j}^{(0,m)}$ follows from the condition (\ref{eqn:background2}). The first few of these functions are: 
	\begin{eqnarray}\label{eqn:metric_k}
		c_{i\:\!j}^{(0,0)}=\delta_{i\:\!j}, \quad c_{i\:\!j}^{(0,1)}=-\frac{1}{2}r^2\,\delta_{i\:\!j}, \quad c_{i\:\!j}^{(0,2)}=\frac{3}{16}r^4\,\delta_{i\:\!j}\\
		\nonumber c_{i\:\!j}^{(0,3)}=-\frac{1}{16}r^6\,\delta_{i\:\!j}, \quad c_{i\:\!j}^{(0,4)}=\frac{5}{256}r^8\,\delta_{i\:\!j}\,.
	\end{eqnarray}
	When $\lambda=0$, the metric functions other than $c_{i\:\!j}^{(0,m)}$ are equal to zero, and the Einstein tensor becomes a function of the curvature parameter only $G^\mu\:\!{}_\nu(0,k)$. One can check by inspection that:
	\begin{eqnarray}\label{eqn:G_k}
		\mathcal{T}_4[G^0\:\!{}_0(0,k)]=-\frac{3(\dot{a}^2+k)}{a^2}\,,\\
		\nonumber \mathcal{T}_4[G^i\:\!{}_j(0,k)]\Big |_{i=j}=-\frac{2a\ddot{a}+\dot{a}^2+k}{a^2}\,,
	\end{eqnarray}
	where the overdot denotes differentiation with respect to time $t$. After inserting these formulas into (\ref{eqn:Einstein}), one can see that the scale factor $a(t)$ satisfies the usual Friedmann equations for a universe filled with dust, with an arbitrary curvature parameter $k$ and the positive cosmological constant $\Lambda$. It is convenient to introduce the Lema\^itre-Hubble parameter $H(t)=\dot{a}(t)/a(t)$, the Lema\^itre-Hubble constant $H_0=H(t_0)$ in which $t_0$ is the age of the universe, and the cosmological parameters $\Omega_m=8\pi\rho(t_0)/(3H_0^2)$ and $\Omega_\Lambda=\Lambda/(3H_0^2)$. Then, the Friedmann equations reduce to one ordinary differential equation for the scale factor:
	\begin{equation}\label{eqn:Friedmann}
		\frac{\ud a}{\ud t}=H_0\,\sqrt{\frac{\Omega_m}{a(t)}+\Omega_\Lambda\,a(t)^2-\frac{k}{H_0^2}}\,,
	\end{equation}
	where the curvature parameter is $k=H_0^2\,\left(\Omega_m+\Omega_\Lambda-1 \right)$. Solutions to an equation of this type are known in terms of the Weierstrass elliptic $\wp$-function (e. g. \cite{1987AJ.....94.1373D}). However, for the purpose of this article, we solve this equation backward in time numerically, with the help of the fourth-order Runge-Kutta method and the initial condition $a(t_0)=1$.
	
	It is worth mentioning that the limit $\lim\limits_{\lambda\rightarrow 0} g_{\mu\:\!\nu}$ provides the FLRW model as a background spacetime for which the amplitude of inhomogeneities vanishes while the spatially flat background $\lim\limits_{(\lambda,k) \rightarrow (0,0)} g_{\mu\:\!\nu}$ is auxiliary and has no physical meaning. Since $a(t)$ satisfies Friedmann's equations with an arbitrarily $k$ and a positive cosmological constant, the quantities $3\dot{a}^2/a^2$ and $-(2a\ddot{a}+\dot{a}^2)/a^2$ do not represent the density and the pressure of any physical fluid.
	
	The main task of this article is to present a method to find the remaining metric functions $c_{i\:\!j}^{(l,m)}$ for $l \geq 1$, so that the approximate Einstein equations (\ref{eqn:Einstein}) up to second order hold within an acceptable uncertainty $|\Delta T^\mu\:\!{}_\nu|$.
	
	\section{Perturbations in the linear order.}\label{Sec:1stOrder}
	When $N=1$, the summation (\ref{eqn:metric_c}) consists of elements $c_{i\:\!j}^{(1,0)}$ and $c_{i\:\!j}^{(0,1)}$. The second one is given by (\ref{eqn:metric_k}), so it remains only $c_{i\:\!j}^{(1,0)}$. Let us introduce only scalar perturbations in the linear order. We restrict the metric elements to the following form:
	\begin{equation}\label{eqn:metric_1}
		c_{i\:\!j}^{(1,0)}=\mathscr{A}_{10}(t)\frac{\partial^2}{\partial x^i \partial x^j}A_{10}(x,y,z)+B_{10}(x,y,z)\,\delta_{i\:\!j}\,.
	\end{equation}
	To complete the construction of the model in the first order, one should determine the two functions of the spatial variables $A_{10}$, $B_{10}$, and the one function of time $\mathscr{A}_{10}$.
	For convenience, we denote the contributions to the Einstein tensor components from the consecutive orders in a way similar to (\ref{eqn:metric_c}):
	\begin{equation}
		\mathcal{T}_N[G^\mu\:\!{}_\nu(\lambda,k)]=\sum\limits_{l=0}^N\sum\limits_{m=0}^{l}[G^\mu\:\!{}_\nu]^{(l-m,m)}\,\lambda^{l-m}\,k^m\,.
	\end{equation}
	The terms $[G^\mu\:\!{}_\nu]^{(0,0)}$ and $[G^\mu\:\!{}_\nu]^{(0,1)}$ are included in (\ref{eqn:G_k}). In the following, we analyze the remaining part $[G^\mu\:\!{}_\nu]^{(1,0)}$.
	
	The assumed form of the metric (\ref{eqn:metric_1}) guarantees that the elements $[G^i\:\!{}_0]^{(1,0)}$ are equal to zero. The other two types of Einstein tensor elements $[G^i\:\!{}_j]^{(1,0)}\big |_{i=j}$ and $[G^i\:\!{}_j]^{(1,0)}\big |_{i\neq j}$ contain non-zero terms with different time dependence. However, one can unify the time dependence of all the terms if only the following differential equation holds:
	\begin{equation}\label{eqn:A1}
		a^2(t)\,\ddot{\mathscr{A}}_{10}(t)+3\,a(t)\,\dot{a}(t)\,\dot{\mathscr{A}}_{10}(t)=\alpha\,,
	\end{equation}   
	where $\alpha$ is a constant. Since the scale factor $a(t)$ is known, this becomes a well-posed equation for the metric function $\mathscr{A}_{10}(t)$. With the auxiliary function $f_1$, such that $f_1(t)=\dot{\mathscr{A}}_{10}(t)$, this equation reduces to the inhomogeneous first-order linear ODE, which can be solved with the variation of constant method. Specifying the initial conditions at some initial time $t_i$, one can write the solution as:
	\begin{eqnarray}\label{eqn:A10}
		\mathscr{A}_{10}(t)=\int\limits_{t_i}^{t}f_1(t')\,\ud t'+\mathcal{C}_1, \quad P_1(t)=3\int\limits_{t_i}^{t}\frac{\dot{a}(t')}{a(t')}\,\ud t'\\
		\nonumber \mathrm{and}\:\:f_1(t)=\left(\mathcal{C}_2+\alpha\int\limits_{t_i}^{t}\frac{\ue^{P_1(t')}}{a^2(t')}\,\ud t' \right)\,\ue^{-P_1(t)}\,.
	\end{eqnarray}
	For simplicity, we may ignore the integration constants $\mathcal{C}_1=0$ and $\mathcal{C}_2=0$. This means that $\mathscr{A}_{10}(t_i)=f_1(t_i)=0$. If $t_i$ is the time of the Big Bang $t_i=0$, then this condition is justifiable. 
	
\begin{figure}[h]
		\centering
		\includegraphics[width=0.45\textwidth]{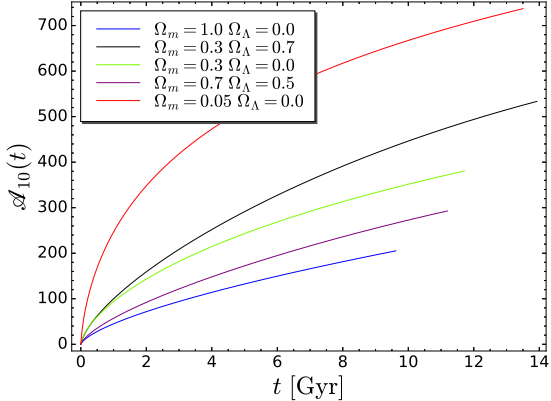}
		\caption{\label{fig:TA} \scriptsize{The metric function $\mathscr{A}_{10}(t)$ calculated for various $\Omega_m$, $\Omega_\Lambda$ parameters. The curves are plotted in the range \mbox{$t\in (t_i,t_0(\Omega_m,\Omega_\Lambda))$}, where $t_i$ is the initial time described in the text and $t_0(\Omega_m,\Omega_\Lambda)$ is the universe age of each cosmological model evaluated for $H_0=67.37\,\mathrm{[km/s/Mpc]}$.}}
	\end{figure}

	In the previous paper \cite{2021EPJC...81..208S}, we investigated perturbations around the Einstein-de Sitter (EdS) background, with $\Omega_m=1$, $\Omega_\Lambda=0$ and $k=0$. The EdS scale factor is $a^{(EdS)}(t)=\mathcal{C}\,t^{2/3}$. The value of the constant $\mathcal{C}=4.85\times 10^{-3}$ can be found using the megaparsec as a unit of length, considering normalization $a^{(EdS)}(t_0)=1$ and calculating the age of the EdS universe $t_0$ for the Lema\^itre-Hubble constant value $H_0=67.37\,\mathrm{[km/s/Mpc]}$ taken from the Planck data \cite{2020A&A...641A...6P}. In the current notation, the respective metric function from the work \cite{2021EPJC...81..208S} is $\mathscr{A}_{10}(t)=t^{2/3}$. It is easy to check that this form of $\mathscr{A}_{10}(t)$ follows from (\ref{eqn:A10}), when $\alpha=\frac{10}{9}\mathcal{C}^2=2.61\times 10^{-5}$. For generic values of the background parameters $(\Omega_m,\Omega_\Lambda)$, the scale factor could no longer have a simple power-law dependence. However, taking the initial time slightly above the Big Bang $t_i=10^{-3}$, the integrals (\ref{eqn:A10}) can be easily evaluated numerically. Figure \ref{fig:TA} shows the resulting metric function $\mathscr{A}_{10}(t)$ for various cosmological parameters $\Omega_m$, $\Omega_\Lambda$, and fixed $\alpha=2.61\times 10^{-5}$.
	
	When the temporal metric function $\mathscr{A}_{10}(t)$ satisfies (\ref{eqn:A10}), the Einstein tensor elements $[G^i\:\!{}_j]^{(1,0)}\big |_{i=j}$ and $[G^i\:\!{}_j]^{(1,0)}\big |_{i\neq j}$ are equal to zero if only the additional condition on the spatial metric functions is satisfied:
	\begin{equation}\label{eqn:A1_B1}
		B_{10}(x,y,z)=\alpha\,A_{10}(x,y,z)\,.
	\end{equation}
	In this way, one can construct a specific dust solution to the linear perturbation theory. 
	
	The energy density $\rho^{(1,0)}=-[G^0\:\!{}_0]^{(1,0)}/(8\pi)$ becomes:
	\begin{equation}\label{eqn:density1}
		\rho^{(1,0)}(t,x,y,z)=\frac{a(t)\,\dot{a}(t)\,\dot{\mathscr{A}}_{10}(t)-\alpha}{8\pi\,a^2(t)}\,\triangle A_{10}(x,y,z)\,.
	\end{equation}
	The symbol $\triangle$ denotes the Laplace operator in Cartesian coordinates.
	The specification of the cosmological parameters $\Omega_m$ and $\Omega_\Lambda$ constrains the time evolution of the first-order density, while its spatial distribution depends on the arbitrary function $A_{10}$.
	
	Since the functions $\mathscr{A}_{10}$ and $B_{10}$ are proportional to $\alpha$, the metric elements $c_{i\:\!j}^{(1,0)}$ are proportional to $\alpha$. From (\ref{eqn:density1}) it follows that the first-order density $\rho^{(1,0)}$ is also proportional to $\alpha$. This means that the value of the constant $\alpha$ does not matter. The value of $\alpha$ can be arbitrarily changed, and then the amplitude $\lambda$ should be redefined so that the first-order metric $\lambda\,c_{i\:\!j}^{(1,0)}$ and the first-order density $\lambda \rho^{(1,0)}$ remain unaffected. For numerical reasons, it is convenient to keep the constant $\alpha$, because it enables one to normalize the function $\mathscr{A}_{10}$. 
	
	\begin{figure}[h]
		\centering
		\includegraphics[width=0.45\textwidth]{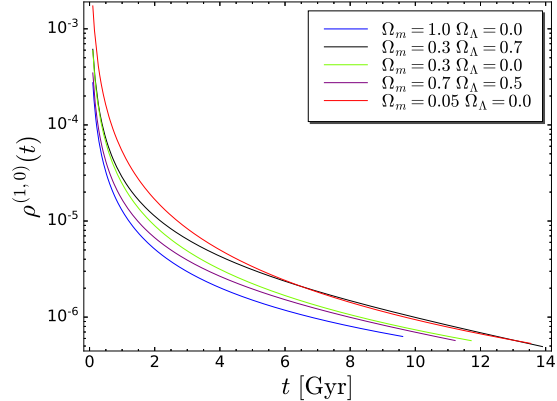}
		\caption{\label{fig:rho1_t} \scriptsize{The time dependence of the first-order contribution to the density. Some specific values of the cosmological parameters $\Omega_m$ and $\Omega_\Lambda$ are considered.}}
	\end{figure} 
	Figure \ref{fig:rho1_t} shows the time dependence of the first-order density $\rho^{(1,0)}(t)=(a(t)\,\dot{a}(t)\,\dot{\mathscr{A}}_{10}(t)-\alpha)/(8\pi\,a^2(t))$. For every value of the background parameters $\Omega_m$ and $\Omega_\Lambda$, the first-order density is a decreasing function of time. The homogeneous density of the FLRW background model $\rho^{(0)}(t)$ is also a decreasing function. If $\rho^{(0)}$ decreases faster than $\rho^{(1,0)}$, the density contrast $\delta=\lambda\,\rho^{(1,0)}/\rho^{(0)}$ may increase with time.
	
	\begin{figure}[h]
		\centering
		\includegraphics[width=0.45\textwidth]{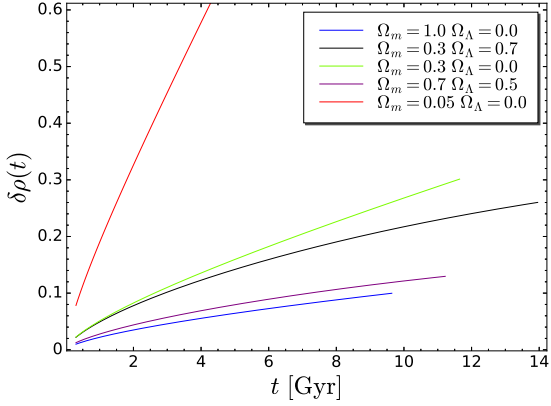}
		\caption{\label{fig:rho_contrast1} \scriptsize{The density contrast $\delta\rho(t)$ for different background FLRW models and the fixed value of the amplitude \mbox{$\lambda=9.68\times 10^{-4}$}.}}
	\end{figure} 
	\begin{figure}[h]
		\centering
		\includegraphics[width=0.45\textwidth]{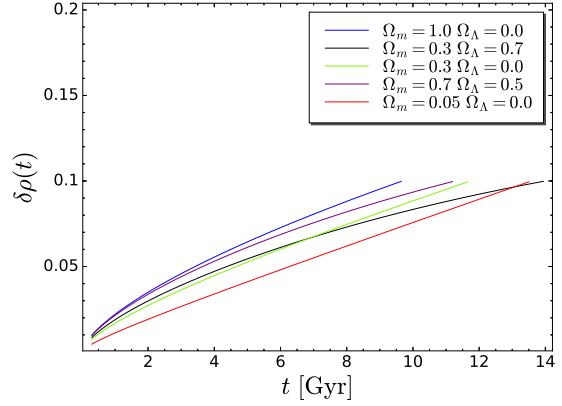}
		\caption{\label{fig:rho_contrast2} \scriptsize{The density contrast $\delta\rho(t)$ evaluated for different values of the amplitude $\lambda$. For the set of the background parameters $(\Omega_m \Omega_\Lambda)\in \{(1,0),(0.3,0.7),(0.3,0),(0.7,0.5),(0.05,0)\}$ the respective values of the amplitude are $\lambda\in\{9.68\times 10^{-4},3.71\times 10^{-4},3.19\times 10^{-4},7.45\times 10^{-4},5.68\times 10^{-5} \}$. }}
	\end{figure} 
	In Fig. \ref{fig:rho_contrast1}, the density contrast is plotted for different cosmological parameters $\Omega_m$, $\Omega_\Lambda$, and the same value of the amplitude $\lambda=9.68\times 10^{-4}$. In Fig. \ref{fig:rho_contrast2}, we show the density contrast for different values of $\lambda$. For each background spacetime, we chose the respective value of $\lambda$ such that the density contrast at the universe age is equal to 0.1. The black and green curves correspond to models that have the same $\Omega_m$ but differ by the value of the cosmological constant. One can observe that the presence of the cosmological constant causes a smaller grouth-rate of the density contrast.
	
	\section{Perturbations in the second order.}\label{Sec:2ndOrder}
	For $N=2$, in the summation (\ref{eqn:metric_c}) appear the elements $c_{i\:\!j}^{(2,0)}$, $c_{i\:\!j}^{(1,1)}$, and $c_{i\:\!j}^{(0,2)}$. The latter is defined by (\ref{eqn:metric_k}), so the elements $c_{i\:\!j}^{(2,0)}$ and $c_{i\:\!j}^{(1,1)}$ remain to be determined. From (\ref{eqn:G_k}) it follows that $[G^\mu\:\!{}_\nu]^{(0,2)}=0$. To achieve the dust solution in second-order perturbation theory, one should analyze contributions to the Einstein tensor elements $[G^\mu\:\!{}_\nu]^{(2,0)}$ and $[G^\mu\:\!{}_\nu]^{(1,1)}$. Equations in the second order become quite tough. To simplify them, let us introduce the following restriction.
	
	As explained in the previous section, the spatial distribution of the density in the first order is given by the arbitrary function $A_{10}$. Let us assume that this function is separable:
	\begin{equation}\label{eqn:Cfunction}
		A_{10}(x,y,z)=C_{10}(x)+C_{10}(y)+C_{10}(z)\,,
	\end{equation}
	where $C_{10}$ is an arbitrary function of one variable. To fulfill (\ref{eqn:A1_B1}) one should also take:
	\begin{equation}
		B_{10}(x,y,z)=D_{10}(x)+D_{10}(y)+D_{10}(z)\,,
	\end{equation}
	with $D_{10}(w)=\alpha\,C_{10}(w)$ and $w=x,y,z$.
	\begin{figure}[h]
		\centering
		\includegraphics[width=0.43\textwidth]{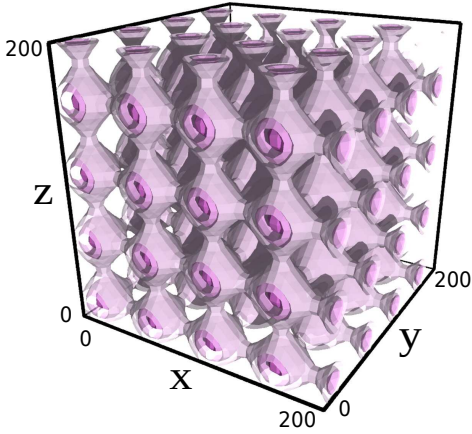}
		\caption{\label{fig:isodensity} \scriptsize{The isodensity surfaces of the first-order density $\rho^{(1,0)}$, for which the function $C_{10}$ is given by (\ref{eqn:C10_sine_comb}), with parameters \mbox{$M=1$}, $P_1=-625/\pi^2$, $Q_1=\pi/25$ and $R_1=0$.}}
	\end{figure} 
	If one gives the arbitrary function $C_{10}$ as a linear combination of the sine functions:
	\begin{equation}\label{eqn:C10_sine_comb}
		C_{10}(w)=\sum\limits_{s=1}^{M}P_s\,\sin(Q_s\,w+R_s)\,,
	\end{equation}
	then the density $\rho^{(1,0)}$ forms an infinite cubic lattice. We plot an exemplary density distribution of this type in Figure \ref{fig:isodensity} for one mode only. Although condition (\ref{eqn:Cfunction}) constrains the space of all possible density distributions, quite complicated distributions are still admissible. The subsequent modes present in (\ref{eqn:C10_sine_comb}) could describe different substructures distributed over the elementary cell. Moreover, such a model is locally inhomogeneous, but for scales much larger than the size of the elementary cell, it becomes homogeneous and isotropic in common sense.
	
	The simplification (\ref{eqn:Cfunction}) eliminates the mixed derivatives that appear in the Einstein tensor elements. After that, elements $[G^i\:\!{}_0]^{(2,0)}$ are proportional to $\alpha\,\dot{\mathscr{A}_{10}}(t)$, elements $[G^i\:\!{}_j]^{(2,0)}\big |_{i\neq j}$ are proportional to $\alpha^2\,a(t)^{-2}$, while diagonal elements $[G^i\:\!{}_j]^{(2,0)}\big |_{i=j}$ comprise three types of terms. The terms of the first type are proportional to $\alpha\,\mathscr{A}_{10}(t)\,a(t)^{-2}$, the terms of the second type are proportional to $\dot{\mathscr{A}_{10}}(t)^2$, and the terms of the third type are proportional to $\alpha^2\,a(t)^{-2}$.
	\begin{figure}[h]
		\centering
		\includegraphics[width=0.45\textwidth]{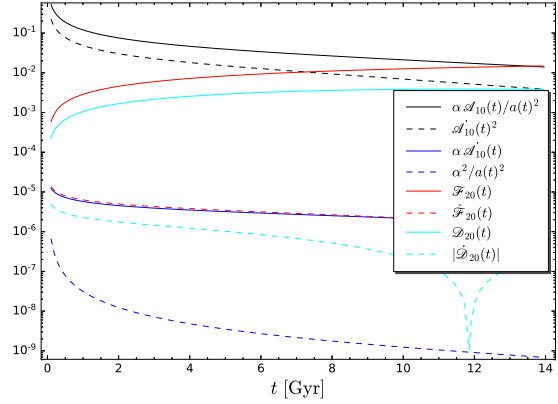}
		\caption{\label{fig:timefunc} \scriptsize{The time dependence of various components of the Einstein tensor elements, calculated for $\Omega_m=0.3$, $\Omega_\Lambda=0.7$ and $\alpha=2.61\times 10^{-5}$. For comparison, the time-dependent metric functions $\mathscr{D}_{20}(t)$, $\mathscr{F}_{20}(t)$, introduced further in the text, and their derivatives.}}
	\end{figure} 
	In Fig. \ref{fig:timefunc} these time-dependent functions are plotted for the scale factor $a(t)$ and the function $\mathscr{A}_{10}(t)$ calculated for the parameters $\Omega_m=0.3$, $\Omega_\Lambda=0.7$ and the already used value $\alpha=2.61\times 10^{-5}$. For other values of the cosmological parameters $\Omega_m$ and $\Omega_\Lambda$ considered in this paper, this figure looks the same. One can observe that for every time $t$ between the Big Bang and the age of the universe, the functions $\alpha\,\dot{\mathscr{A}_{10}}(t)$ and $\alpha^2\,a(t)^{-2}$ are at least five orders of magnitude smaller than $\alpha\,\mathscr{A}_{10}(t)\,a(t)^{-2}$ and $\dot{\mathscr{A}_{10}}(t)^2$. The energy-momentum tensor elements consist of the time-dependent functions mentioned above multiplied by some terms that contain spatial metric functions and their derivatives. If the frequency of the sine functions in (\ref{eqn:C10_sine_comb}) is not too high, then the derivatives of the spatial metric functions are relatively small, and all the $[G^\mu\:\!{}_\nu]^{(2,0)}$ terms proportional to $\alpha\,\dot{\mathscr{A}_{10}}(t)$ and $\alpha^2\,a(t)^{-2}$ could be neglected.
	
	The remaining two types of terms present in $[G^i\:\!{}_j]^{(2,0)}\big |_{i=j}$ can be eliminated if the metric functions in the order $(2,0)$ are the following:
	\begin{eqnarray}
		c_{i\:\!j}^{(2,0)}=\left(\mathscr{A}_{20}(t)C_{20}(x^i)+\mathscr{D}_{20}(t)\sum\limits_{s=1}^{3}D_{20}(x^s)\right)\delta_{i\:\!j}+ \\
		\nonumber+\mathscr{F}_{20}(t)\,F_{20}(x^i,x^j)\,(1-\delta_{i\:\!j})\,.
	\end{eqnarray}
	The two of the time-dependent functions are given \mbox{explicitly}:
	\begin{equation}\label{eqn:D20}
		\mathscr{D}_{20}(t)=a^2(t)\,\dot{\mathscr{A}}_{10}(t)^2\,, 
	\end{equation}
	\begin{equation}
		\mathscr{F}_{20}(t)=\frac{1}{4}a^2(t)\,\dot{\mathscr{A}}_{10}(t)^2+\alpha\,\mathscr{A}_{10}(t)\,.
	\end{equation}
	The third one should satisfy the differential equation:
	\begin{equation}\label{eqn:A2}
		a^2(t)\,\ddot{\mathscr{A}}_{20}(t)+3\,a(t)\,\dot{a}(t)\,\dot{\mathscr{A}}_{20}(t)=2\,\alpha\,\mathscr{A}_{10}(t)\,.
	\end{equation}
	This equation is similar to (\ref{eqn:A1}). Since the function $\mathscr{A}_{10}$ is known from the first order, it can be integrated numerically with the help of the same method. 
	
	The spatial metric functions should satisfy the following equations:
	\begin{equation}
		D''_{20}(w)=-\frac{1}{2}\,C''_{10}(w)^2\,,
	\end{equation}
	\begin{equation}
		\frac{\partial^2}{\partial w\partial v}F_{20}(v,w)=-C''_{10}(v)\,C''_{10}(w)\,,
	\end{equation}
	\begin{equation}\label{eqn:C20}
		C_{20}(w)=-\frac{1}{4}\,C'_{10}(w)\,C'''_{10}(w)\,,
	\end{equation}
	where the prime denotes the derivative with respect to one of the spatial variables $u,v\in\{x,y,z\}$. For the function $C_{10}$ given by (\ref{eqn:C10_sine_comb}), it is easy to find the explicit solution to these equations:
	\begin{eqnarray}
		D_{20}(w)\!=\!-\frac{1}{8}\sum\limits_{s=1}^{M}P_s^2Q_s^4\!\left(w^2+\frac{\cos(2Q_s w\!+\!2R_s)}{2Q_s^2}\right)+ \\
		\nonumber+\frac{1}{4}\sum\limits_{s=1}^{M}\sum\limits_{\substack{r=1 \\ r\neq s}}^{M} P_s P_r Q_s^2 Q_r^2\!\left[\frac{\cos((Q_s\!-\!Q_r)w+R_s\!-\!R_r)}{(Q_s\!-\!Q_r)^2}- \right. \\
		\nonumber\left.-\frac{\cos((Q_s\!+\!Q_r)w+R_s\!+\!R_r)}{(Q_s\!+\!Q_r)^2} \right]\,,
	\end{eqnarray}
	\begin{equation}
		F_{20}(v,w)\!=\!-\!\sum\limits_{s=1}^{M}\sum\limits_{r=1}^{M}\!\!P_s  P_r Q_s Q_r\!\cos(Q_s v+R_s)\cos(Q_r w+R_r)\,,
	\end{equation}
	\begin{equation}
		C_{20}(w)\!=\!\frac{1}{4}\sum\limits_{s=1}^{M}\sum\limits_{r=1}^{M}P_s  P_r Q_s Q_r^3 \cos(Q_s v +R_s)\cos(Q_r w+R_r)\,.
	\end{equation}
	Integration constants are ignored for simplicity.
	
	For the metric functions described above, the leading-order terms proportional to $\alpha\,\mathscr{A}_{10}(t)\,a(t)^{-2}$ and $\dot{\mathscr{A}_{10}}(t)^2$ are eliminated from $[G^i\:\!{}_j]^{(2,0)}\big |_{i=j}$. Instead, some terms proportional to the derivatives of the metric functions $\dot{\mathscr{D}}_{20}$, $\dot{\mathscr{F}}_{20}$, and $\ddot{\mathscr{F}}_{20}$ appear in the other elements of $[G^\mu\:\!{}_\nu]^{(2,0)}$. The function $\dot{\mathscr{D}}_{20}$ changes sign, but as could be seen in Fig. \ref{fig:timefunc}, its absolute value $|\dot{\mathscr{D}}_{20}|$ remains very small compared to the leading-order terms. Similarly, the derivatives $\dot{\mathscr{F}}_{20}$ and $\ddot{\mathscr{F}}_{20}$ are also small and could be neglected. In effect, all the Einstein tensor elements $[G^\mu\:\!{}_\nu]^{(2,0)}\approx 0$, apart from the second-order contribution to the energy density $\rho^{(2,0)}=-[G^0\:\!{}_0]^{(2,0)}/(8\pi)$.
	With the help of equations (\ref{eqn:D20}-\ref{eqn:C20}) the energy density could be simplified to the following form:
	\begin{equation}
		\rho^{(2,0)}(t,x,y,z)=H(t,x)+H(t,y)+H(t,z)\,,
	\end{equation}
	where
	\begin{eqnarray}\label{eqn:rho2_time}
		H(t,w)=\left[\frac{\alpha\mathscr{A}_{10}(t)}{2a^2(t)}-\frac{\dot{\mathscr{A}}_{20}(t)\dot{a}(t)}{4a(t)}\right]C'_{10}(w)C'''_{10}(w)+\\
		\nonumber+\left[\frac{\dot{\mathscr{A}}_{10}(t)^2}{2}+\frac{\alpha\mathscr{A}_{10}(t)}{a^2(t)}-\frac{\mathscr{A}_{10}(t)\dot{\mathscr{A}_{10}(t)}\dot{a}(t)}{a(t)}\right]C''_{10}(w)^2- \\
		\nonumber-\frac{\alpha\dot{\mathscr{A}}_{10}(t)\dot{a}(t)}{a(t)}C_{10}(w)C''_{10}(w)+\frac{3\alpha\dot{a}(t)}{a(t)}\frac{\ud}{\ud t}\left[a^2\dot{\mathscr{A}}_{10}^2 \right]C_{10}(w)
	\end{eqnarray}
	Each term of $H(t,w)$ represents a decreasing function of time. The last two are the smallest.
	
	Similarly, for the order $(1,1)$, one can choose the same proposition for the structure of the metric functions:
	\begin{eqnarray}
		c_{i\:\!j}^{(1,1)}=\left(\mathscr{A}_{11}(t)C_{11}(x^i)+\mathscr{D}_{11}(t)\sum\limits_{s=1}^{3}D_{11}(x^s)\right)\delta_{i\:\!j}+ \\
		\nonumber+\mathscr{F}_{11}(t)\,F_{11}(x^i,x^j)\,(1-\delta_{i\:\!j})\,.
	\end{eqnarray}
	The Einstein tensor elements $[G^i\:\!{}_j]^{(1,1)}\big |_{i=j}=0$ and $[G^i\:\!{}_0]^{(1,1)}=0$ vanish if the metric functions satisfy the following relations. The two of the time-dependent functions are equal to the first-order function $\mathscr{A}_{10}$:
	\begin{equation}
		\mathscr{D}_{11}(t)=\mathscr{F}_{11}(t)=\mathscr{A}_{10}(t)\,,
	\end{equation} 
	while the third one should satisfy the differential equation:
	\begin{equation}
		a^2(t)\,\ddot{\mathscr{A}}_{11}(t)+3\,a(t)\,\dot{a}(t)\,\dot{\mathscr{A}}_{11}(t)=\mathscr{A}_{10}(t)\,.
	\end{equation}
	Again, this equation is of the same type as (\ref{eqn:A1}) and (\ref{eqn:A2}); therefore, it can be integrated numerically in a similar manner.
	The spatial metric functions are given by:
	\begin{equation}
		D_{11}(w)=C_{10}(w)-\frac{1}{2}\,w\,C'_{10}(w)\,,
	\end{equation}
	\begin{equation}
		F_{11}(v,w)=\frac{1}{2}\left(v\,C'_{10}(w)+w\,C'_{10}(v) \right)\,,
	\end{equation}
	\begin{equation}
		C_{11}(w)=C''_{10}(w)\,.
	\end{equation}
	After that, the only non-zero components of the Einstein tensor in order $(1,1)$ are $[G^0\:\!{}_0]^{(1,1)}$ and $[G^i\:\!{}_j]^{(1,1)}\big |_{i\neq j}$. However, the components $[G^i\:\!{}_j]^{(1,1)}\big |_{i\neq j}$ are proportional to $\alpha/a^2(t)$ and can be neglected. The density of order $(1,1)$ $\rho^{(1,1)}=-[G^0\:\!{}_0]^{(1,1)}/(8\pi)$ can be simplified to the following form:
	\begin{equation}\label{eqn:rho11}
		\rho^{(1,1)}(t,x,y,z)=K(t,x,r)+K(t,y,r)+K(t,z,r)\,,
	\end{equation}
	where
	\begin{eqnarray}
		K(t,w,r)=\left[\frac{\dot{\mathscr{A}}_{10}(t)\dot{a}(t)}{2a(t)}\,r^2-\frac{\mathscr{A}_{10}(t)}{a^2(t)}\right]C''_{10}(w)+\\\nonumber+\ddot{\mathscr{A}}_{10}(t)\left[\frac{1}{2}w\,C'_{10}(w)-C_{10}(w) \right]+\frac{\dot{\mathscr{A}}_{11}(t)\dot{a}(t)}{a(t)}C_{10}(w)\,. 
	\end{eqnarray}
	Once again, this point shows that the model presented here is adequate only for the description of the local neighborhood of the selected observer, where the coordinate distance $r^2=x^2+y^2+z^2$ is small.
	
	\subsection{The convergence}
	To construct an approximate solution to the Einstein equations, we used perturbative methods and neglected some terms identified as small compared to the leading-order terms. It is then necessary to verify whether the solution converges to a universe filled with dust. If the model converges properly, then the uncertainty of the energy-momentum tensor $\Delta T^\mu\:\!{}_\nu$ should decrease when the contributions to the metric tensor from the consecutive orders are taken into account. 
	
\begin{figure}[h!]
		\centering
		\includegraphics[width=0.43\textwidth]{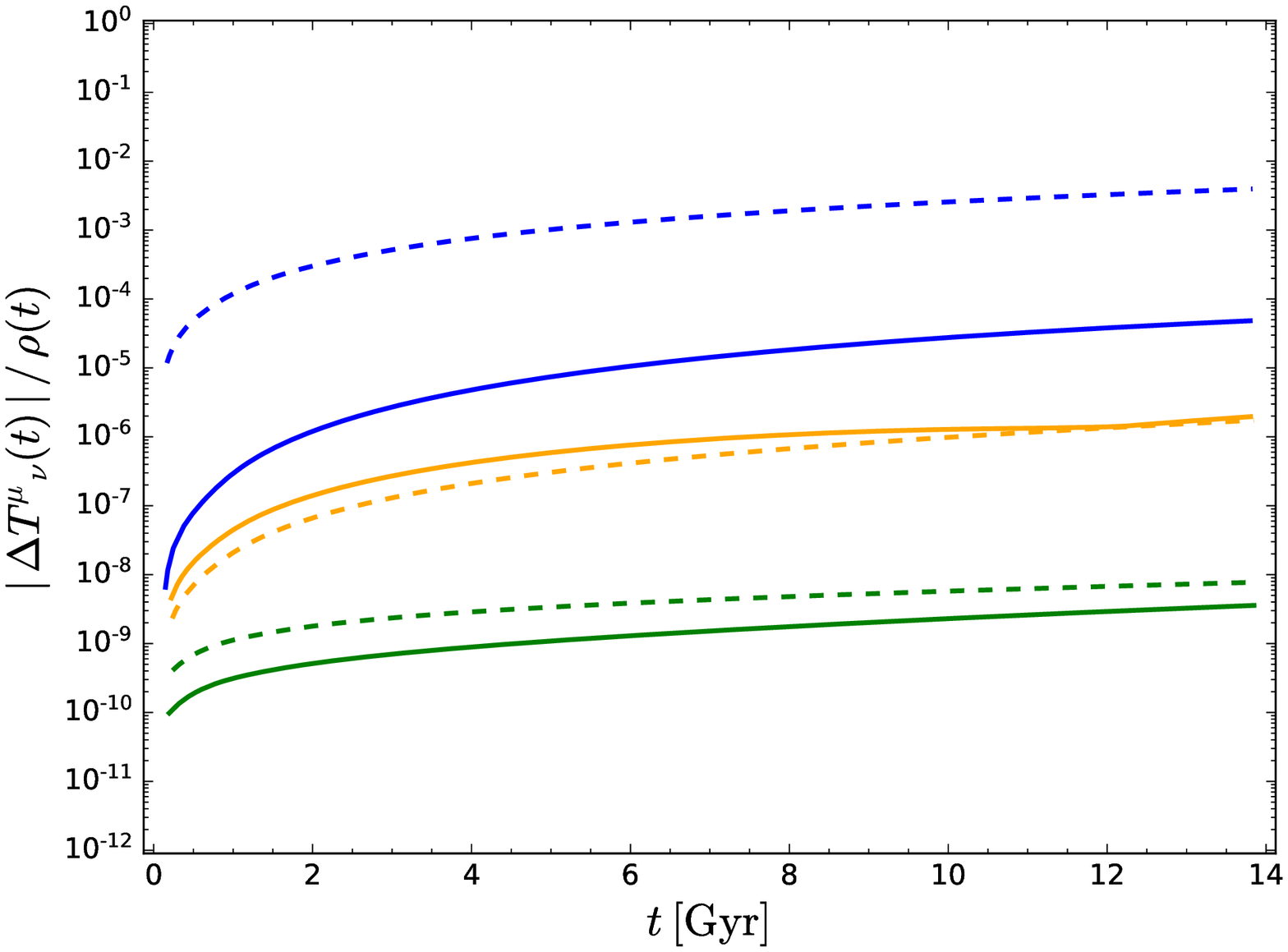}
		\includegraphics[width=0.43\textwidth]{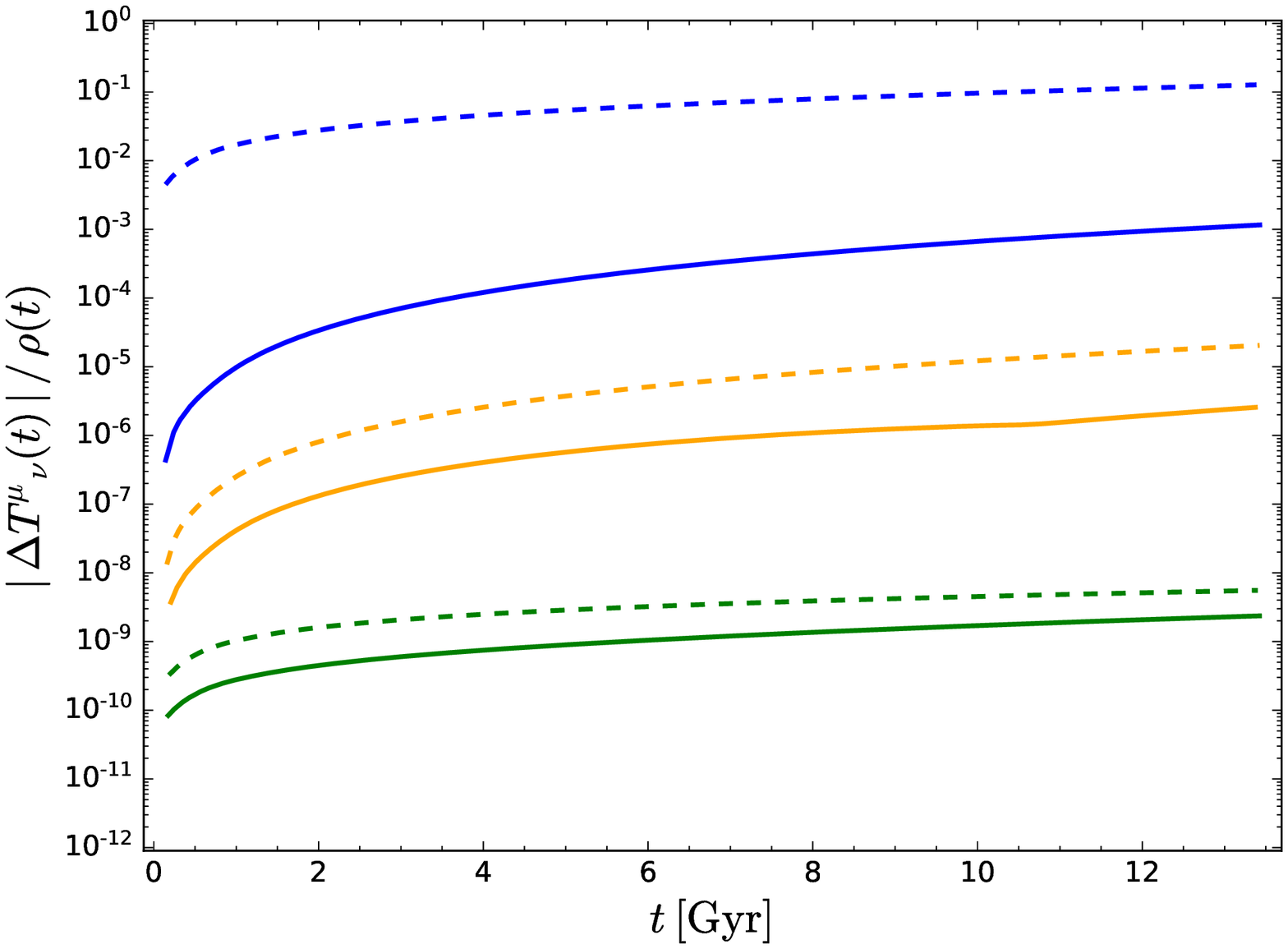}
		\includegraphics[width=0.43\textwidth]{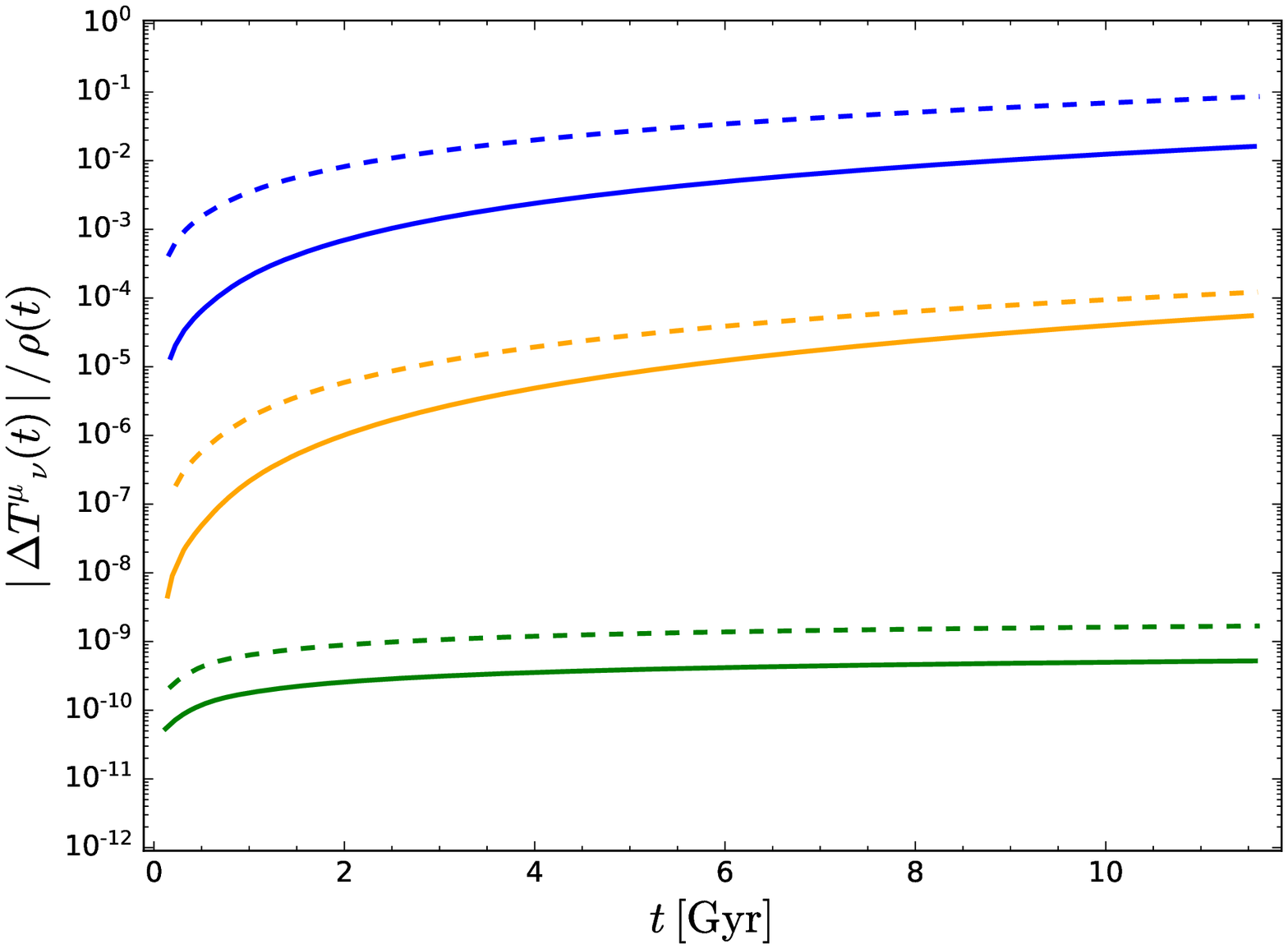}
		\caption{\label{fig:convergence1} \scriptsize{The maximum of $|\,\Delta T^\mu\:\!{}_\nu\,(t)|/\rho(t)$ over the elementary cell. The \emph{top} graph corresponds to $(\Omega_m,\Omega_\Lambda)=(0.3,0.7)$, the \emph{middle} is done for $(\Omega_m,\Omega_\Lambda)=(0.3,0.6)$, and the \emph{bottom} - $(\Omega_m,\Omega_\Lambda)=(0.3,0.0)$. The \emph{blue} curves describe $\Delta T^i\:\!{}_j \big |_{i=j}$, the \emph{orange} curves - $\Delta T^i\:\!{}_0$, and the \emph{green} curves - $\Delta T^i\:\!{}_j \big |_{i\neq j}$. \emph{Dashed} lines are generated for the linear order metric, while the \emph{solid} lines correspond to the metric up to the second order.}}
	\end{figure}

	In Fig. \ref{fig:convergence1} we plot the components $\Delta T^\mu\:\!{}_\nu$ in units of energy density $\rho$ for the Taylor expansion of the Einstein tensor $\mathcal{T}_N[G^\mu\:\!{}_\nu(\lambda,k)]$ evaluated up to the third order \mbox{$N=3$}. We take the arbitrary function $A_{10}$ as the one mode sine function for which the density distribution is plotted in Fig. \ref{fig:isodensity}. The value of the amplitude $\lambda$ is such that the first-order density $\rho^{(1,0)}$ evaluated at the maximum of the overdensity equals $0.025$ in the critical density units. In each plot, we show the maximum of the absolute value $|\,\Delta T^\mu\:\!{}_\nu\,|$ among the values at one hundred randomly chosen points within the elementary cell. The blue curves represent pressure-like terms $\Delta T^i\:\!{}_j \big |_{i=j}$, the orange curves correspond to $\Delta T^i\:\!{}_0$, and the green curves describe $\Delta T^i\:\!{}_j \big |_{i\neq j}$. The results for the metric truncated at linear order are plotted by dashed lines. When the metric functions are considered up to the second order, the corresponding results are plotted by solid lines.
	
	In each case, pressure-like terms give the greatest contribution to the discrepancy $\Delta T^\mu\:\!{}_\nu$ between the energy-momentum tensor of the model and the energy-momentum tensor of dust. When the second-order metric is taken into account, the absolute value of the pressure-like terms is smaller than in the case of linear-order perturbations. In the case of the spatially flat universe, with $\Omega_m=0.3$, $\Omega_\Lambda=0.7$, and $k=0$, the second-order results are improved by more than two orders of magnitude compared to the first-order predictions. The value of the pressure-like terms $\Delta T^i\:\!{}_j \big |_{i=j}/\rho$ below $10^{-5}$ is acceptable, since the pressure generated by the proper motions of members of a typical galaxy cluster is of the order of $10^{-6}\,\rho$.
	For a non-flat background, the value of the pressure-like terms is higher. We observe the values of $\Delta T^i\:\!{}_j \big |_{i=j}/\rho$ around $10^{-4}$ and $10^{-3}$ for the parameters $(\Omega_m,\Omega_\Lambda)=(0.3,0.6)$ and $(\Omega_m,\Omega_\Lambda)=(0.3,0.0)$, respectively. In these cases, to achieve better accuracy, third-order corrections to the metric tensor are needed.
	
	In each plot, the terms $\Delta T^i\:\!{}_0$ and $\Delta T^i\:\!{}_j \big |_{i\neq j}$ are improved when the second-order metric is taken into account or their values do not change too much. Generally, we may conclude that the convergence of the presented model is satisfactory.
	
	\subsection{The gauge}\label{Sec: gauge}
\begin{figure}[h]
		\centering
		\includegraphics[width=0.45\textwidth]{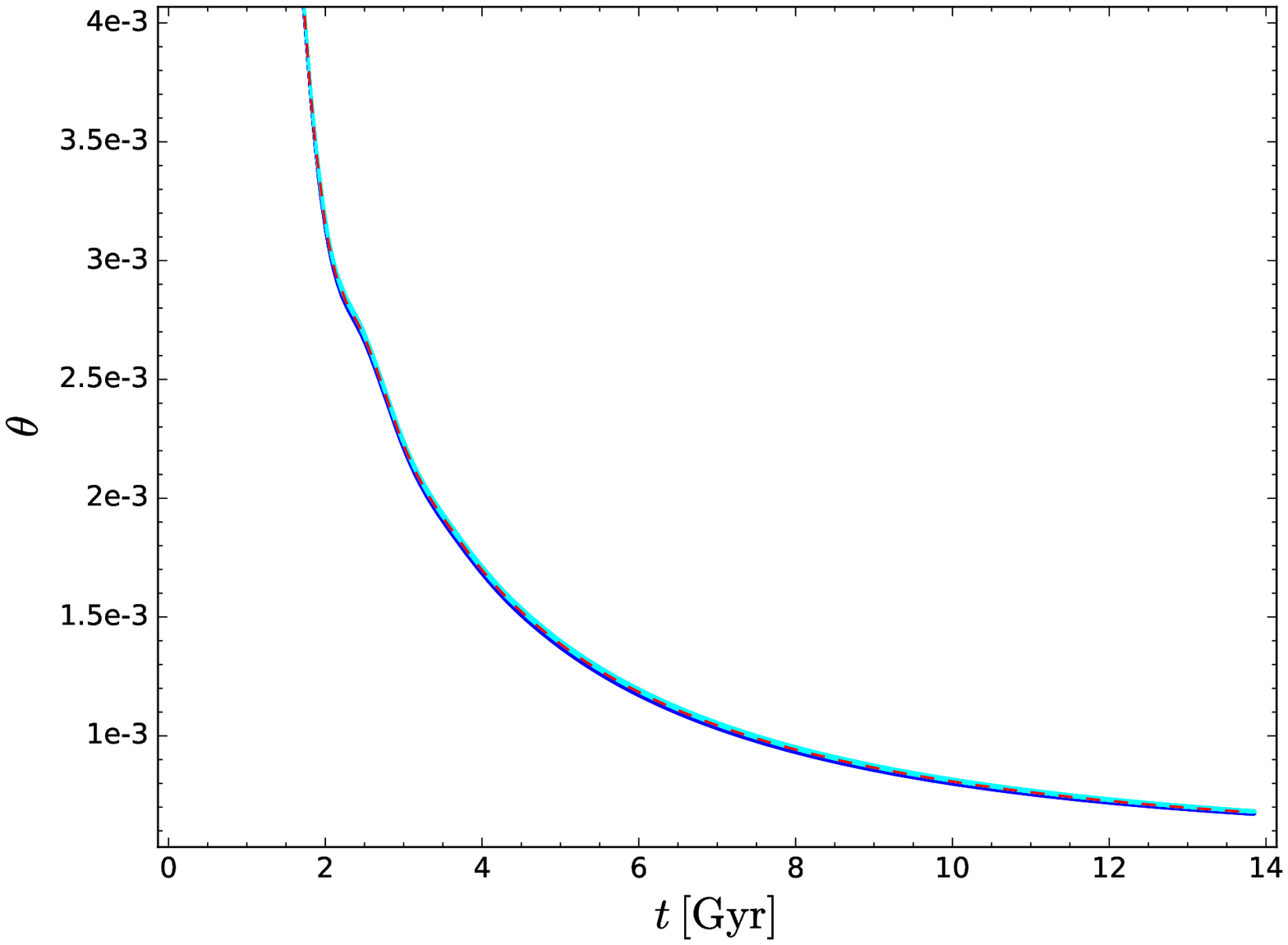}
		\includegraphics[width=0.45\textwidth]{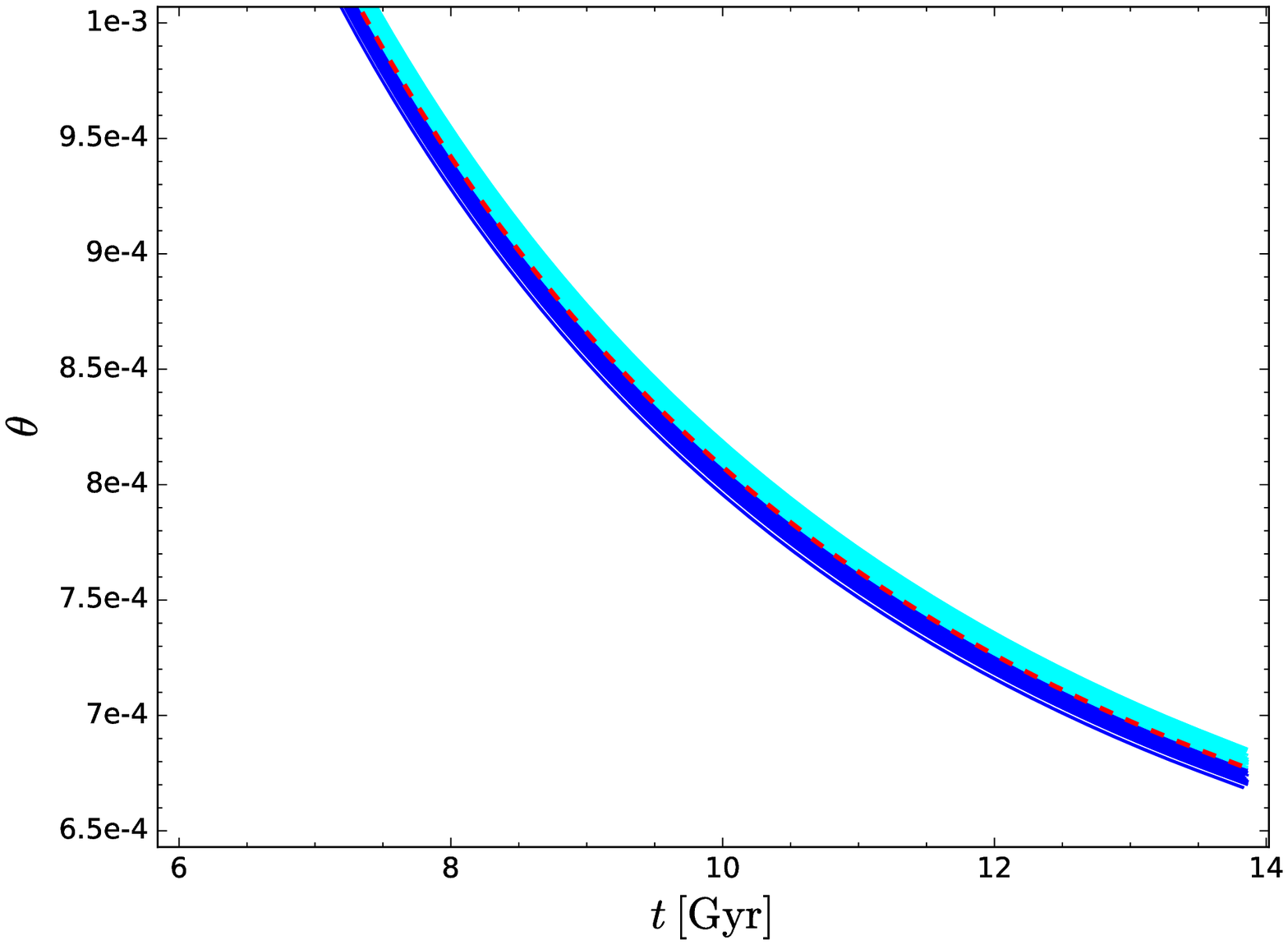}
		\caption{\label{fig:theta} \scriptsize{The expansion scalar $\theta$ as a function of time for an exemplary model on the background $(\Omega_m,\Omega_\Lambda)=(0.3,0.7)$. The expansion scalar evaluated in the overdensity regions is plotted by the \emph{blue} curves, while $\theta$ in the underdensity regions are drawn by the \emph{light blue} curves. The \emph{dashed red} curve represents the expansion scalar in the background space-time. The axes range in the \emph{bottom} panel helps us to visualize more details. }}
	\end{figure}

	In Fig. \ref{fig:theta} we show the expansion scalar $\theta=\nabla_\mu\,U^\mu$ for an exemplary model. We take the background parameters $(\Omega_m,\Omega_\Lambda)=(0.3,0.7)$, fix the arbitrary function $A_{10}$ as the one mode sine function defining the density distribution presented in Fig. \ref{fig:isodensity}, and take the value of the amplitude $\lambda$ so that the first-order density $\rho^{(1,0)}$ at the maximum of the overdensity is $0.025$ in the critical density units. Then, we randomly generate one hundred points within the elementary cell. The time dependence of $\theta$ at the points whose density is higher than the background density $\Omega_m=0.3$ is plotted in Fig. \ref{fig:theta} by the blue curves. The relation $\theta(t)$ at the points where the model density is lower than $0.3$ is plotted by the light blue curves. By the red dashed curve, we show the time dependence of the expansion scalar of the background. It is seen that in the considered inhomogeneous model the space is expanding everywhere, and the rate of expansion is slightly changed compared to the background. The overdense regions expand a bit slower than the underdense areas. Because the bundle of time-like geodesics is not collapsing, the world-lines of initially separated observers would not cross with each other, and the Gaussian normal coordinates (\ref{eqn:metric_GaussianNormal}) cover the whole space-time. 
	
	The perturbation theory is carried out within the synchronous comoving gauge when the form of the metric tensor (\ref{eqn:metric_GaussianNormal}) is fixed, so that $g_{0\:\!0}=-1$, $g_{0\:\! i}=0$, and the velocity field $U^\mu$ is unperturbed. These conditions do not completely determine the gauge. There is the following gauge freedom left:
	\begin{equation}\label{eqn:gauge_trans}
		x^i \mapsto \chi^i(x^j)\,,
	\end{equation}
	where $\chi^i$, for $i=1,2,3$, are three arbitrary functions of the spatial variables only. The gauge transformation (\ref{eqn:gauge_trans}) leaves the form of the metric tensor (\ref{eqn:metric_GaussianNormal}) unchanged and the velocity field unperturbed. It is necessary to check whether the considered perturbations are not fictitious. The phenomena which are not physical can be removed by a gauge transformation.
	
	Let us specify the following transformation:
	\begin{equation}
		\chi^i(x^j)=x^i+\lambda\,[\chi^{(1,0)}]^i(x^j)+k\,\,[\chi^{(0,1)}]^i(x^j)\,,
	\end{equation}
	where $\chi^{(1,0)}$ and $\chi^{(0,1)}$ can be chosen arbitrarily. The model density is transformed under this gauge transformation in the following way:
	\begin{multline}\label{eqn:gauge1}
		\rho(t,\chi^i(x^j))\approx \rho(t,x^j)+\lambda\,\frac{\partial \rho}{\partial x^m}\,[\chi^{(1,0)}]^m(x^j)+ \\
		+k\,\frac{\partial \rho}{\partial x^m}\,[\chi^{(0,1)}]^m(x^j)\,.
	\end{multline}
	If the derivative $\partial\rho/\partial x^j$ is non-zero, then spurious density fluctuations can be introduced in each order of perturbation theory. The background spacetime is homogeneous, so its density $\rho_0$ does not depend on the spatial coordinates $x^j$. Then it is not possible to generate the first-order density perturbation $\rho^{(1,0)}$ by a gauge transformation. The first-order density is not fictitious.
	
	Suppose that we have the following density:
	\begin{equation}
		\rho(t,x^j)=\rho_0(t)+\lambda\,\rho^{(1,0)}(t,x^j)\,.
	\end{equation}
	Since it depends on the spatial variables, the gauge transformation (\ref{eqn:gauge1}) introduces non-physical density in the second order:
	\begin{equation}
		\widetilde{\rho}^{(2,0)}(t,x^j)=\frac{\partial\rho^{(1,0)}}{\partial x^m}\,[\chi^{(1,0)}]^m(x^j)\,,
	\end{equation}
	and
	\begin{equation}
		\widetilde{\rho}^{(1,1)}(t,x^j)=\frac{\partial\rho^{(1,0)}}{\partial x^m}\,[\chi^{(0,1)}]^m(x^j)\,.
	\end{equation}
	However, since the gauge functions $\chi^i$ depend only on spatial variables, the time dependence of $\widetilde{\rho}^{(2,0)}$ and $\widetilde{\rho}^{(1,1)}$ is inherited from the first-order density $\rho^{(1,0)}$. The time dependence of $\widetilde{\rho}^{(2,0)}$ differs from the time dependence of every term in (\ref{eqn:rho2_time}). As a result, the second-order density $\rho^{(2,0)}$ is also physical and cannot be produced from $\rho^{(1,0)}$ by a gauge transformation. The same reasoning shows that the density $\rho^{(1,1)}$, given by the formula (\ref{eqn:rho11}), is also physical. In this way, analysis of the time dependence of the density distribution enables one to remove the gauge freedom.
	
	\section{Conclusions.}
	In this work we present a specific solution to the second-order cosmological perturbation theory around a generic FLRW model filled with dust. The non-zero spatial curvature of the background and a positive cosmological constant are allowed. We simplified the metric tensor so that only one arbitrary function $C_{10}$ ramains and defines the spatial distribution of the density, while the other metric functions are determined from the model construction. If $C_{10}$ is a periodic function, for example a linear combination of the sine functions, then the density forms an infinite cubic lattice. The distribution of the substructures within the elementary cell depends on the choice of $C_{10}$. Although the density contrast could be an increasing function of time, the space of the model universe is expanding at each point. For that reason, the model does not intend to describe the formation of individual structures but gives a tool for handling differences between large areas occupied by the galaxy clusters and the cosmic voids. Since the metric tensor is explicitly given and has a relatively simple form, the model can be used to examine quantitatively how these differences affect cosmological observables.

	\bibliography{SikoraGlod2021}
	\bibliographystyle{unsrt}

\end{document}